\documentclass{aa}  

\usepackage{graphicx}
\usepackage{txfonts}

\begin{document}

\title{The white-dwarf cooling sequence of NGC~6791: a unique tool for
       stellar evolution}

\author{Enrique Garc\'{\i}a--Berro\inst{1,2},
        Santiago Torres\inst{1,2}, 
        Isabel Renedo\inst{1,2},
        Judit Camacho\inst{1,2},
        Leandro G. Althaus\inst{3,4},        
        Alejandro H. C\'orsico\inst{3,4},
        Maurizio Salaris\inst{5}
        \and
        Jordi Isern\inst{2,6}}
\institute{Departament de F\'\i sica Aplicada,
           Universitat Polit\`ecnica de Catalunya,
           c/Esteve Terrades 5, 
           08860 Castelldefels, Spain\
           \and       
           Institute for Space  Studies of Catalonia,
           c/Gran Capit\`a 2--4, Edif. Nexus 104,   
           08034  Barcelona, Spain\ 
           \and
           Facultad de Ciencias Astron\'omicas y Geof\'{\i}sicas, 
           Universidad Nacional de La Plata, 
           Paseo del Bosque s/n, 1900 La Plata, Argentina\
           \and
           Instituto de Astrof\'{\i}sica de La Plata (CCT La Plata), 
           CONICET, 1900 La Plata, Argentina\
           \and
           Astrophysics Research Institute, 
           Liverpool John Moores University, 
           12 Quays House, 
           Birkenhead, CH41 1LD, 
           United Kingdom\
           \and           
           Institut de Ci\`encies de l'Espai, CSIC, 
           Campus UAB, Facultat de Ci\`encies, Torre C-5, 
           08193 Bellaterra, Spain}

\offprints{E. Garc\'\i a--Berro}

\date{\today}

\abstract{NGC 6791 is a  well-studied, metal-rich open cluster that it
          is so  close to us that  can be imaged  down to luminosities
          fainter  than that  of the  termination of  its  white-dwarf
          cooling sequence, thus allowing for an in-depth study of its
          white dwarf population.}
         {White dwarfs  carry important information  about the history
          of  the cluster.   We  use observations  of the  white-dwarf
          cooling  sequence to constrain  important properties  of the
          cluster  stellar  population, such  as  the  existence of  a
          putative population of massive helium-core white dwarfs, and
          the properties  of a  large population of  unresolved binary
          white dwarfs.   We also investigate the use  of white dwarfs
          to disclose  the presence  of cluster subpopulations  with a
          different  initial chemical  composition, and  we  obtain an
          upper  bound  to the  fraction  of hydrogen-deficient  white
          dwarfs.}
         {We  use  a Monte  Carlo  simulator  that employs  up-to-date
          evolutionary  cooling   sequences  for  white   dwarfs  with
          hydrogen-rich   and  hydrogen-deficient   atmospheres,  with
          carbon-oxygen and  helium cores.  The  cooling sequences for
          carbon-oxygen  cores account  for the  delays  introduced by
          both  $^{22}$Ne sedimentation  in  the liquid  phase and  by
          carbon-oxygen phase separation upon crystallization.}
         {We  do  not find  evidence  for  a  substantial fraction  of
           helium-core white dwarfs, and hence our results support the
           suggestion  that  the origin  of  the  bright  peak of  the
           white-dwarf luminosity function can only be attributed to a
           population  of unresolved  binary white  dwarfs.  Moreover,
           our  results indicate  that if  this hypothesis  is  at the
           origin  of  the bright  peak,  the  number distribution  of
           secondary masses  of the population  of unresolved binaries
           has  to increase  with  increasing mass  ratio between  the
           secondary and primary  components of the progenitor system.
           We also find that  the observed cooling sequence appears to
           be   able   to  constrain   the   presence  of   progenitor
           subpopulations with different chemical compositions and the
           fraction of hydrogen-deficient white dwarfs.}
         {Our simulations  place interesting constraints  on important
          characteristics of  the stellar populations  of NGC~6791. In
          particular, we find that  the fraction of single helium-core
          white dwarfs must be  smaller than 5\%, that a subpopulation
          of stars with zero metallicity  must be $\la 12$\%, while if
          the adopted  metallicity of  the subpopulation is  solar the
          upper limit is  $\sim 8$\%.  Finally, we also  find that the
          fraction   of  hydrogen-deficient   white  dwarfs   in  this
          particular cluster is surprinsingly small ($\la 6\%$).}
\keywords{stars:  white  dwarfs --  stars:  luminosity function,  mass
          function  --  (Galaxy:)   open  clusters  and  associations:
          general  --   (Galaxy:)  open  clusters   and  associations:
          individual: NGC 6791}

\titlerunning{The white dwarf cooling sequence of NGC~6791}
\authorrunning{E. Garc\'\i a--Berro et al.}

\maketitle


\section{Introduction}
\label{intro}

White dwarfs are the  most common stellar evolutionary endpoint. Their
evolution can be described as a very slow cooling process.  Given that
these fossil  stars are abundant  and long-lived objects,  they convey
important  information  about  the  formation  and  evolution  of  all
Galactic populations (Hansen \&  Liebert 2003; Althaus et al.  2010a).
In particular, white dwarfs can  be employed as reliable cosmic clocks
to infer  the ages of a  wide variety of stellar  populations, such as
the Galactic disk and halo  (Winget et al.  1987; Garc\'\i a--Berro et
al. 1988a,b; Isern  et al.  1998) and the system  of globular and open
clusters (Kalirai  at al.  2001; Hansen et  al.  2002).  Additionally,
they can be  used to place constraints on  exotic elementary particles
(Isern et al.  1992; C\'orsico et  al.  2001; Isern et al. 2008) or on
alternative theories  of gravitation  (Garc\'\i a--Berro et  al. 1995;
Benvenuto et al. 2004; Garc\'\i a--Berro et al. 2011).

The use of white dwarfs as cosmochronometers relies on the accuracy of
the  theoretical cooling  sequences.  Recently,  Garc\'\i  a--Berro et
al. (2010)  have demonstrated  that the slow  down of the  white dwarf
cooling  rate  owing  to  the  release of  gravitational  energy  from
$^{22}$Ne sedimentation  (Bravo et al.   1992; Bildsten \&  Hall 2001;
Deloye  \&  Bildsten 2002)  and  carbon-oxygen  phase separation  upon
crystallization  (Garc\'\i a--Berro  et al.   1988a; Segretain  et al.
1994) is of fundamental importance to reconcile the age discrepancy of
the  very old,  metal-rich  open cluster  NGC~6791.   This raises  the
possibility  of using  the  white-dwarf  luminosity  function of  this
cluster to constrain its fundamental properties.

NGC~6791  is one  of the  oldest Galactic  open clusters  --  see, for
instance,  Bedin  et al.   (2005)  and  Kalirai  et al.   (2007),  and
references therein -- and  it is so close to us that  it can be imaged
down to  very faint luminosities.   A deep luminosity function  of its
well populated white dwarf  sequence has been recently determined from
HST observations (Bedin et al.  2008a) and displays a sharp cut-off at
low  luminosities, caused by  the finite  age of  the cluster,  plus a
secondary  peak at  larger  luminosities, most  likely  produced by  a
population of  unresolved binary white  dwarfs (Bedin et  al.  2008b).
These characteristics  make this cluster  a primary target to  use the
white dwarf cooling sequence to constrain the presence of a population
of  massive helium-core  white dwarfs,  the properties  of  the binary
white   dwarf  population,  the   hypothetical  presence   of  cluster
subpopulations   of  different  metallicity,   and  the   fraction  of
hydrogen-deficient  (non-DA)  white  dwarfs.   Here we  address  these
issues by  means of  Monte Carlo based  techniques aimed  at producing
synthetic  color-magnitude  diagrams   and  luminosity  functions  for
NGC~6791 white  dwarfs, which can  be compared with  the observational
data.

Firstly, we investigate in detail  the nature of the secondary peak of
the white-dwarf luminosity function.  This peak has been atributted to
a population of unresolved binary white dwarfs (Bedin et al. 2008b) or
to the  existence of a  population of single helium-core  white dwarfs
(Hansen  2005).  This  is  a  crucial question  because  if the  first
hypothesis is true,  the amplitude of the secondary  peak is such that
leads to an unusual fraction  of binary white dwarfs, thus challenging
our understanding of the physical processes that rule the formation of
binary white dwarfs,  especially at high metallicities.  Consequently,
we also study other explanations for the existence of a secondary peak
in  the  white-dwarf luminosity  function,  like  the  existence of  a
population of  single helium-core white dwarfs.   This explanation was
first  put forward by  Hansen (2005)  and Kalirai  et al.   (2007) and
later  was challenged,  among  others,  by van  Loon  et al.   (2008).
However,  as  we  will  show  below, this  explanation  results  in  a
white-dwarf luminosity function that is at odds with the observed one,
and hence the most likely explanation for the secondary bright peak is
the  population of  unresolved binaries.   Indeed, there  is  a strong
reason to suspect  that the bright peak of  the white-dwarf luminosity
function of  NGC~6791 is caused  by a population of  unresolved binary
white dwarfs, namely, that the two peaks of the white-dwarf luminosity
function are separated by $0.75^{\rm  mag}$. This is just exactly what
it  should be  expected if  the bright  peak is  caused by  equal mass
binaries.   Hence, if  this explanation  for  the bright  peak of  the
white-dwarf luminosity function  is true this, in turn,  enables us to
study the  properties of the  population of such binary  white dwarfs.
Specifically, we study how different distributions of secondary masses
of  the  unresolved binary  white  dwarfs  affect the  color-magnitude
diagram and the white-dwarf luminosity function.

Additionally, we test whether  the white-dwarf luminosity function can
provide   an  independent   way  to   check  for   the   existence  of
subpopulations  within  a  stellar  system.   The  presence  of  these
subpopulations has been found in several Galactic globular clusters of
which $\omega$ Cen is,  perhaps, the most representative one (Calamida
et al.   2009).  The  appearance of NGC~6791  color-magnitude diagram,
and the lack of any  significant chemical abundance spread (Carraro et
al.     2006)    points   toward    a    very   homogeneous    stellar
population. However,  a recent paper  by Twarog et al.   (2011) raises
the possibility  that there  could be a  1~Gyr age  difference between
inner  and outer  regions of  the cluster.   Nevertheless,  within the
field covered by  the white dwarf photometry used  in our analysis the
age difference should  be negigible.  This is different  from the case
of  individual  Galactic globular  clusters  that  host  at least  two
distinct  populations with  approximately the  same age  and different
abundance  patterns of  the C-N-O-Na  elements  -- see  Milone et  al.
(2010) for   a  recent   brief  review.    Taking  advantage   of  the
well-populated  cooling  sequence   in  the  observed  color-magnitude
diagram,   we  test   whether   modeling  the   cluster  white   dwarf
color-magnitude  alone  can  exclude  the presence  of  subpopulations
generated  by progenitors with  a metallicity  different from  the one
measured spectroscopically.

As a final study, we use the luminosity function of NGC~6791 (Bedin et
al.   2008a),  and  the  fact  that white  dwarfs  with  hydrogen-rich
atmospheres  (of  the DA  type)  and non-DA  white  dwarfs  cool at  a
different rate, to place constraints on the fraction of these objects.
This is  an important point because  Kalirai et al.  (2005) have found
that in  the young  open star  cluster NGC~2099 there  is a  dearth of
non-DA  white dwarfs.   These authors  attributed the  lack  of non-DA
objects to  the fact  that possibly the  hot, high-mass  cluster white
dwarfs  -- white  dwarfs  in this  cluster  are estimated  to be  more
massive than  field objects  in the same  temperature range --  do not
develop sufficiently extended helium  convection zones to allow helium
to be brought to the surface and turn a hydrogen-rich white dwarf into
a helium-rich one.  Studying the  fraction of non-DA white dwarfs in a
different  open cluster  could  provide additional  insight into  this
question.  Moreover, Kalirai et al.  (2007) analyzed a sample of $\sim
15$ white  dwarfs in NGC  6791, and although  the sample was  far from
being complete, all them were of the DA type.

The  paper  is organized  as  follows.   Sect.~2  summarizes the  main
ingredients of our  Monte Carlo code plus the  other basic assumptions
and procedures necessary to  evaluate the characteristics of the white
dwarf   population    for   the   different    simulations   presented
here. Specifically, we discuss  the most important ingredients used to
construct  reliable  color-magnitude  diagrams and  the  corresponding
white-dwarf luminosity functions. Sect.~3 presents the main results of
our  simulations for  all points  already mentioned  in  this section.
Finally, Sect.~4 closes the paper with a summary of our conclusions.


\section{Modeling NGC~6791}

\subsection{The Monte-Carlo simulator}
\label{MC}

The  photometric  properties  of   NGC~6791  were  simulated  using  a
Monte-Carlo technique.  Our Monte  Carlo simulator has been previously
employed  to  model the  Galactic  disk  and  halo field  white  dwarf
populations  (Garc\'\i a--Berro  et al.   1999; Torres  et  al.  2002;
Garc\'\i a--Berro  et al. 2004),  with excellent results.   We briefly
describe  here  the  most  important  ingredients  of  our  simulator.
Synthetic  main-sequence  stars  are  randomly drawn  according  to  a
Salpeter-like initial mass function that in the mass range relevant to
NGC~6791 white dwarf progenitors  ($M> 1.0\, M_{\sun}$) is essentially
identical to the ``universal'' initial mass function of Kroupa (2001),
and a burst of star formation starting 8 Gyr ago, lasting 0.1~Gyr (the
exact  value is  not critical  for  our analysis).   If not  otherwise
stated,  we account  for a  population of  unresolved  detached binary
white dwarfs adopting a  fraction $f_{\rm bin}=0.54$ of binary systems
in the  main sequence, which leads  to a fraction  of unresolved white
dwarf binaries of 36\%.  In our fiducial model we use the distribution
of secondary masses previously employed  by Bedin et al.  (2008b), but
other  distributions  were   also  used  (see  Sect.~\ref{binary})  to
constrain  the  properties  of   the  binary  population.   We  employ
main-sequence  lifetimes from  the calculations  by Weiss  \& Ferguson
(2009) for   $Z=0.04$,   $Y=0.325$   models,   which   correspond   to
[Fe/H]$\sim$0.4,   a   metallicity    consistent   with   the   recent
spectroscopic  determinations  of Carraro  et  al.~(2006), Gratton  et
al.~(2006)   and  Origlia   et  al.~(2006).    For  the   white  dwarf
initial-final  mass  relationship we  used  that  of  Ferrario et  al.
(2005),  which  is  appropriate  for metal-rich  stars,  although  our
results  are  fairly  insensitive   to  the  precise  choice  of  this
function.   For   instance,    when   the   up-to-date   semiempirical
initial-final mass function of Catal\'an et al. (2008) is adopted, the
results  are almost  indistinguishable  from those  obtained with  the
former relationship.

Given the cluster age, the time  at which each synthetic star was born
and its associated main sequence  lifetime, we kept track of the stars
able to  evolve to  the white dwarf  stage, and we  interpolated their
colors  and  luminosities  using  the  theoretical  cooling  sequences
described in the following  subsection.  For unresolved binary systems
we performed the  same calculation for the secondary  and we added the
fluxes and computed the  corresponding colors.  The photometric errors
were drawn according to Gaussian distributions, using the Box-M\"uller
algorithm  as  described  in  Press  et  al.   (1986).   The  standard
photometric  errors in magnitude  and color  were assumed  to increase
linearly  with the magnitude  following Bedin  et al.   (2005, 2008a).
Finally,  we  added  the  distance modulus  of  NGC~6791,  $(m-M)_{\rm
  F606W}=13.44$, and  its color excess  $E($F606W-F814W$)=0.14$ (Bedin
et  al.~2008a,b) to  obtain  a synthetic  white dwarf  color-magnitude
diagram,  and  from this  we  computed  the corresponding  white-dwarf
luminosity function. The distance modulus adopted here agrees with the
recent estimate of Grundahl et al.~(2008), which is based on a cluster
eclipsing binary system.

\subsection{The cooling sequences}
\label{cooling}

The  cooling sequences  adopted  in this  work  were those  previously
employed by  Garc\'\i a--Berro et al.  (2010),  which were extensively
discussed  in Althaus et  al.  (2010b).   These cooling  sequences are
appropriate for white dwarfs  with hydrogen-rich atmospheres, and were
computed  from stellar models  with the  metallicity of  NGC~6791.  In
summary, these  sequences were obtained from star  models with stellar
masses at the ZAMS ranging from  1 to $3\, M_{\sun}$ and were followed
through the thermally  pulsing and mass-loss phases on  the AGB to the
white  dwarf stage.   The resulting  white dwarf  masses  were 0.5249,
0.5701, 0.593,  0.6096, 0.6323,  0.6598 and 0.7051  $M_{\sun}$.  Along
the white dwarf cooling track  the calculations include the release of
gravitational energy resulting from  $^{22}$Ne diffusion in the liquid
phase   and  from  phase   separation  of   carbon  and   oxygen  upon
crystallization.   The  energy  contributions  of these  two  physical
separation   processes  were   computed  by   closely   following  the
prescriptions of  Isern et al.  (1997)  and Isern et  al.  (2000).  We
stress here that  this was made in a  self-consistent way, because the
energy  release  is  locally  coupled  to  the  equations  of  stellar
evolution.   Finally,  we  mention   that  our  calculations  rely  on
realistic  boundary conditions  for  cool white  dwarfs,  as given  by
non-gray model atmospheres (Rohrmann  et al.  2010).  For non-DA white
dwarfs a new  set of cooling sequences was computed  from the same set
of progenitors, using  the same physical inputs as  adopted in Althaus
et al.  (2010b), the only difference being the chemical composition of
the atmosphere,  for which we  adopted pure helium.  Details  of these
calculations will be  presented elsewhere.  The bolometric corrections
and color transformations for this set of cooling sequences were those
of Bergeron et al. (1995).


\section{Results}

\subsection{A population of single helium-core white dwarfs?}
\label{heliof}

\begin{figure*}[t]
\vspace{12cm}    
\includegraphics{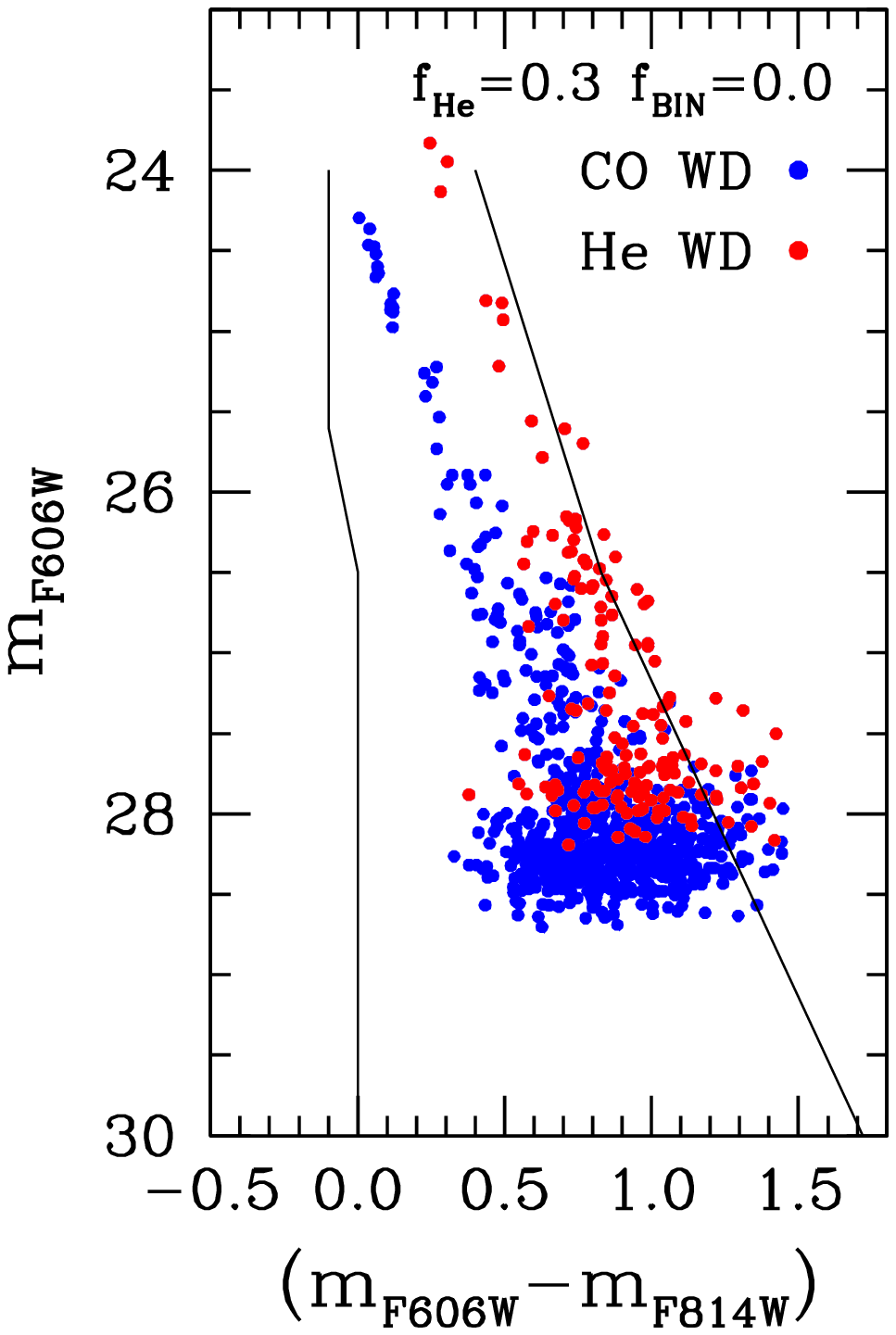}
\includegraphics{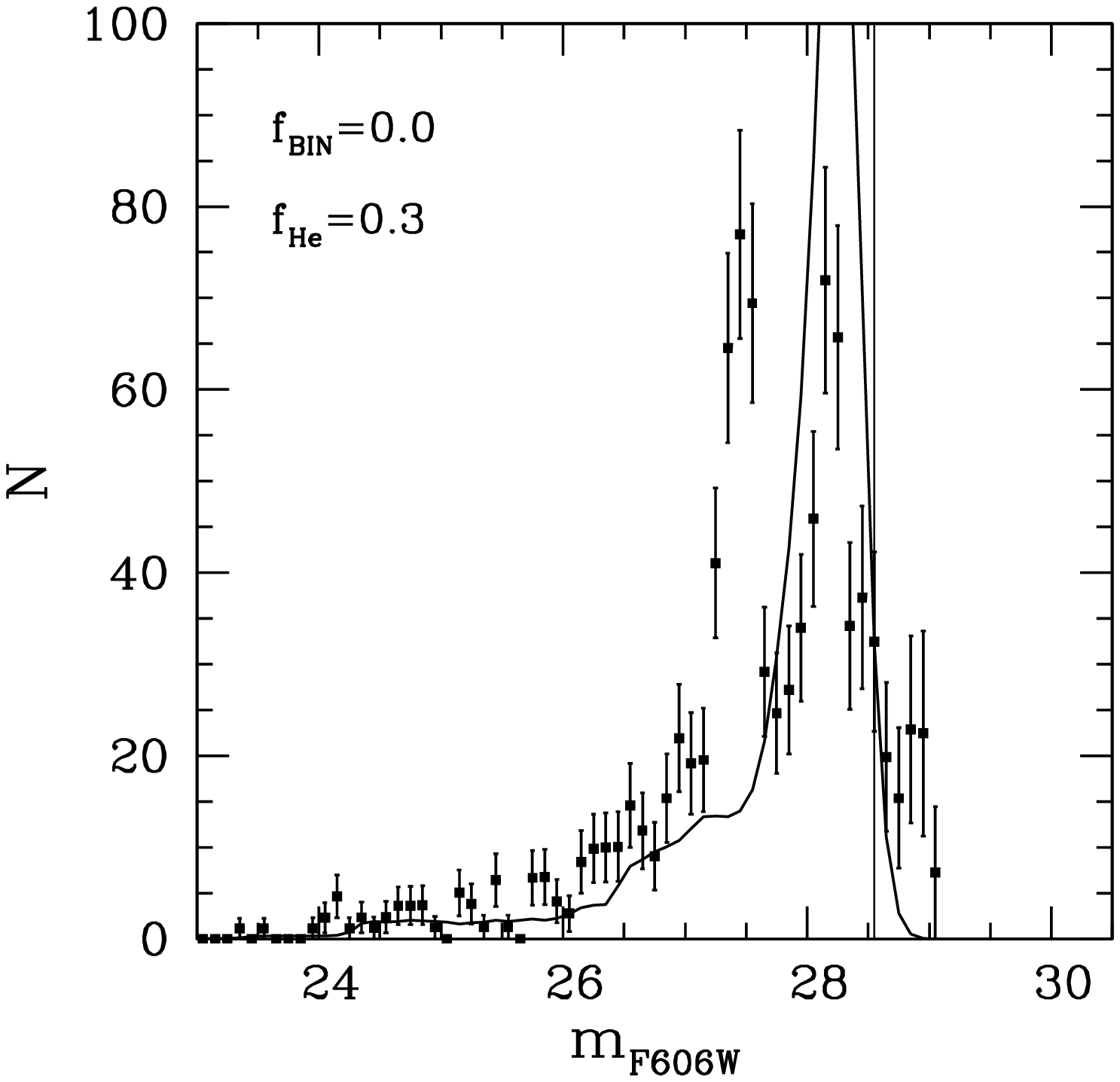}
\includegraphics{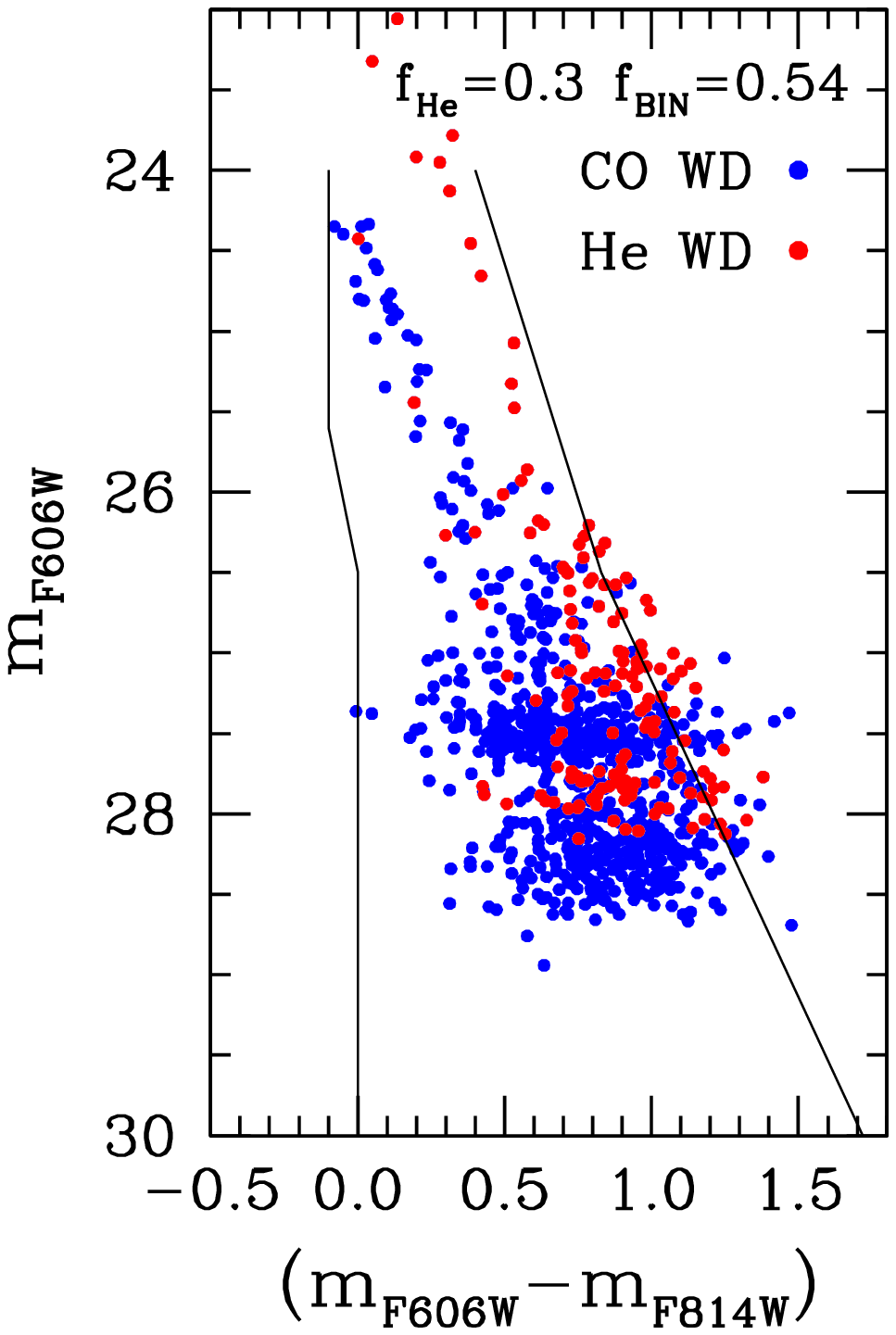} 
\includegraphics{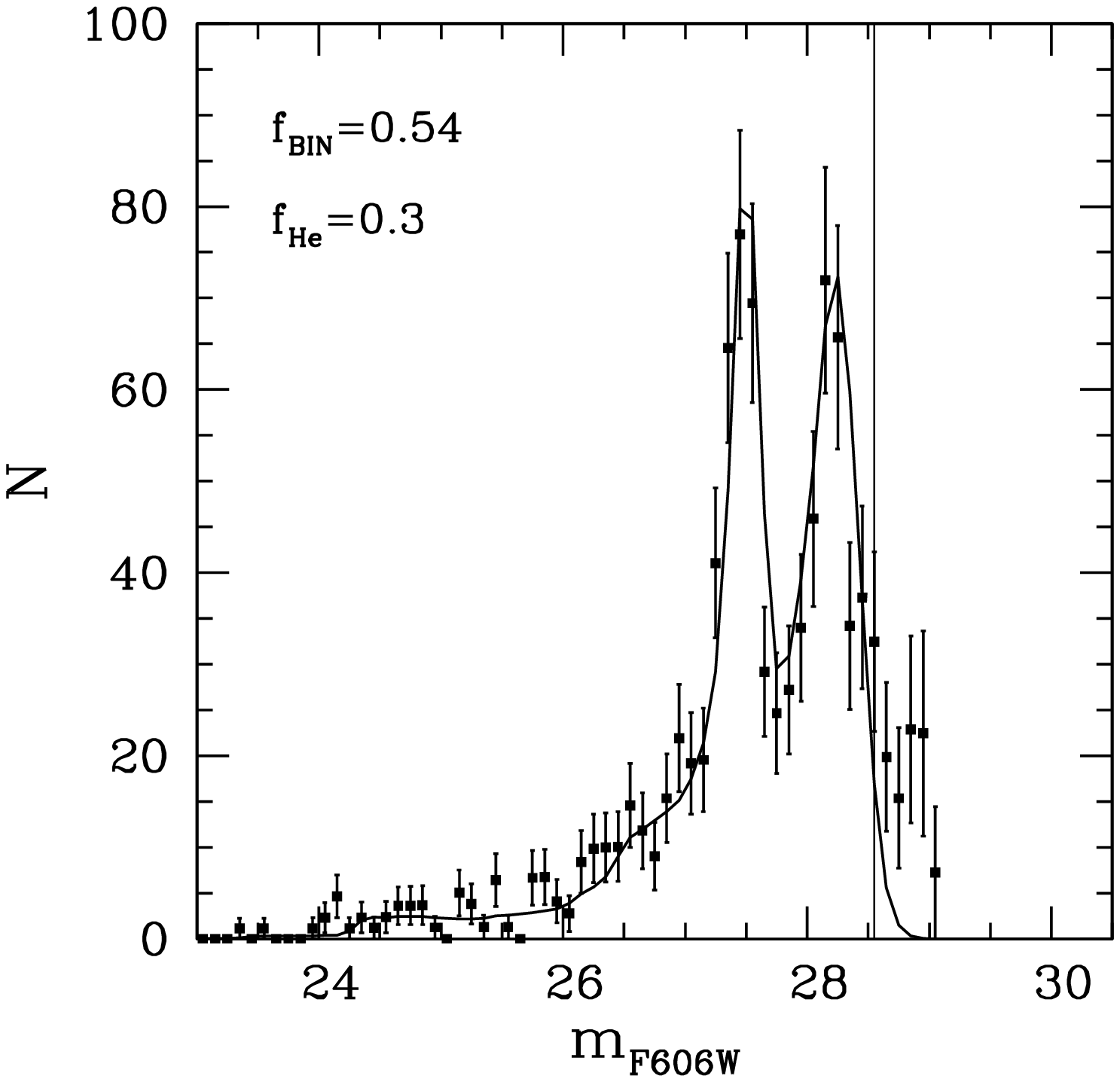} 
 
\caption{Color-magnitude diagrams  (left   panels)  of  the  synthetic
         population of  carbon-oxygen white dwarfs  (blue symbols) and
         of   helium-core   white  dwarfs   (red   symbols)  and   the
         corresponding   white-dwarf   luminosity   functions   (right
         panels).    The   observational   selection   area   in   the
         color-magnitude diagram of Bedin et al.  (2005) is also shown
         using  thin lines.  These  boundaries in  the color-magnitude
         exclude low-mass helium-core white dwarfs.  The observational
         white-dwarf  luminosity function  of Bedin  et  al.  (2008b),
         which was corrected for  incompleteness, is shown using black
         squares.  Each theoretical luminosity function corresponds to
         an average of $10^4$  Monte Carlo realizations.  The vertical
         thin line  marks the magnitude limit  $\simeq 28.5^{\rm mag}$
         above which  the completeness  level of the  photometry falls
         below 50\%.  The  top panels correspond to the  case in which
         $f_{\rm He}=0.3$ and $f_{\rm bin}=0.0$ are adopted, while for
         the bottom ones we adopted $f_{\rm He}=0.3$ and $f_{\rm
         bin}=0.54$.   See the  online edition  of the  journal  for a
         color version of this figure.}
\label{helio}
\end{figure*}

As mentioned,  the bright peak of the  white-dwarf luminosity function
of NGC~6791 has been attributed  either to a huge population of binary
white  dwarfs (Bedin  2008b) or  to the  existence of  popopulation of
single  helium-core white  dwarfs (Hansen  2005; Kalirai  2007).  This
would  indicate that  at the  very  high metallicity  of this  cluster
mass-loss  at  the  tip of  the  red  giant  branch would  be  largely
enhanced.   This  possibility has  been  recently investigated,  among
others, by Meng et al.  (2008),  who found -- on the basis of specific
assumptions  about the  minimum  envelope mass  of  red or  asymptotic
giant-branch stars  -- that  for $Z\ga 0.02$  helium white  dwarfs are
likely the result  of the evolution of stars  with masses smaller than
$1.0\ \,  M_{\sun}$.  However, Prada  Moroni \& Straniero  (2009) have
demonstrated that  when the star  loses the envelope and  departs from
the  red giant  branch with  a core  mass slightly  smaller  than that
required for  helium ignition, a non-negligible possibility  of a late
helium  ignition  exists,  and  low-mass carbon-oxygen  white  dwarfs,
rather  than  helium-core  white  dwarfs, are  produced.   Thus,  more
studies to  constrain a hypothetical population  of single helium-core
white dwarfs are needed.

To  this end  we have  proceeded as  follows.  At  the  metallicity of
NGC~6791, the  helium-core mass at the  helium flash is  $\sim 0.45 \,
M_{\sun}$,  practically   constant  with  initial  mass,   up  to  the
transition  to non-degenerate  cores (Weiss  \& Ferguson  2009). Also,
adopting an  age of 8~Gyr, the mass  at the turn-off is  $\sim 1.15 \,
M_{\sun}$ and the  maximum mass of progenitors that  could have made a
helium-core white dwarf is $\sim 1.8\, M_{\sun}$.  This means that the
range of  masses of possible  progenitors of single white  dwarfs with
helium  cores  is  between  $\sim  1.15$ and  $\sim  1.8\,  M_{\sun}$.
Accordingly,  we draw  a  pseudo-random  number for  the  mass of  the
progenitor  using our  initial mass  function and  we consider  that a
fraction $f_{\rm He}$ of stars  between 1.15 and $2.0\, M_{\sun}$ have
lost the  envelope near the  tip of the  red giant branch  and produce
helium-core white  dwarfs with masses  0.2 and $0.5\,  M_{\sun}$.  The
adopted cooling  sequences for helium-core  white dwarfs are  those of
Althaus et al. (2009).

The  results of our  simulations are  displayed in  the top  panels of
Fig.~\ref{helio}, where we  show the synthetic color-magnitude diagram
and white-dwarf  luminosity function  of the cluster  for the  case in
which we  adopt $f_{\rm He}=0.3$  and $f_{\rm bin}=0$.   Clearly, this
scenario is  unable to  reproduce the bright  peak of the  white-dwarf
luminosity function and the corresponding clump in the color-magnitude
diagram.   Indeed, most helium-core  white dwarfs  are located  in the
same region of the color-magnitude diagram where regular carbon-oxygen
white dwarfs  are placed, the only  difference is the  position of the
cut-off.   In  fact,  helium-core   white  dwarfs  pile-up  at  higher
luminosities than carbon-oxygen ones, as expected, but at luminosities
slightly fainter than that of  the observational bright peak.  The net
result  of this  is that  the population  of helium-core  white dwarfs
partially  overlaps with  the  peak produced  by normal  carbon-oxygen
white  dwarfs and,  consequently, the  faint peak  of the  white-dwarf
luminosity function broadens.

\begin{figure*}[t]
\vspace{12cm}    
\includegraphics{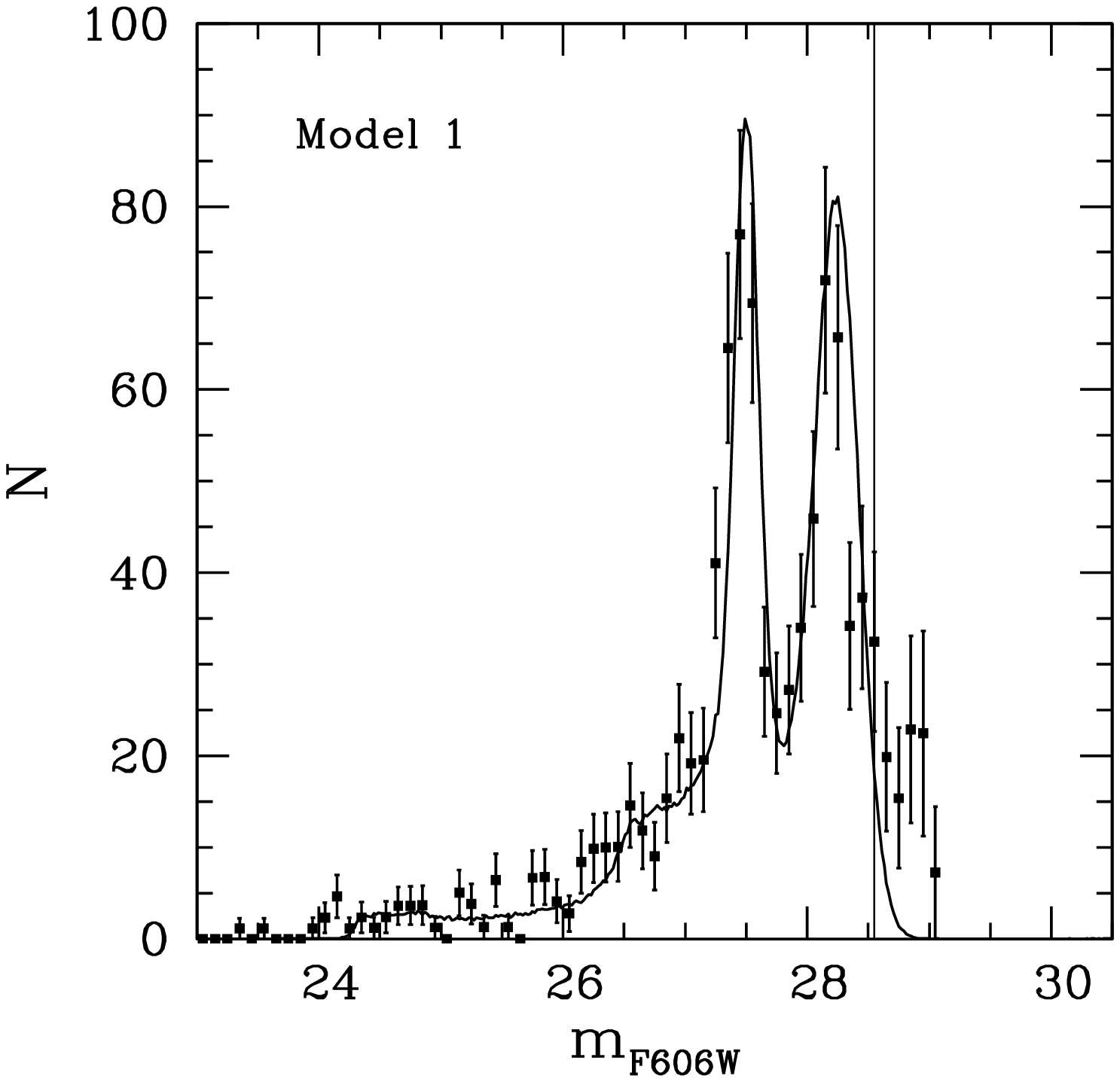}
\includegraphics{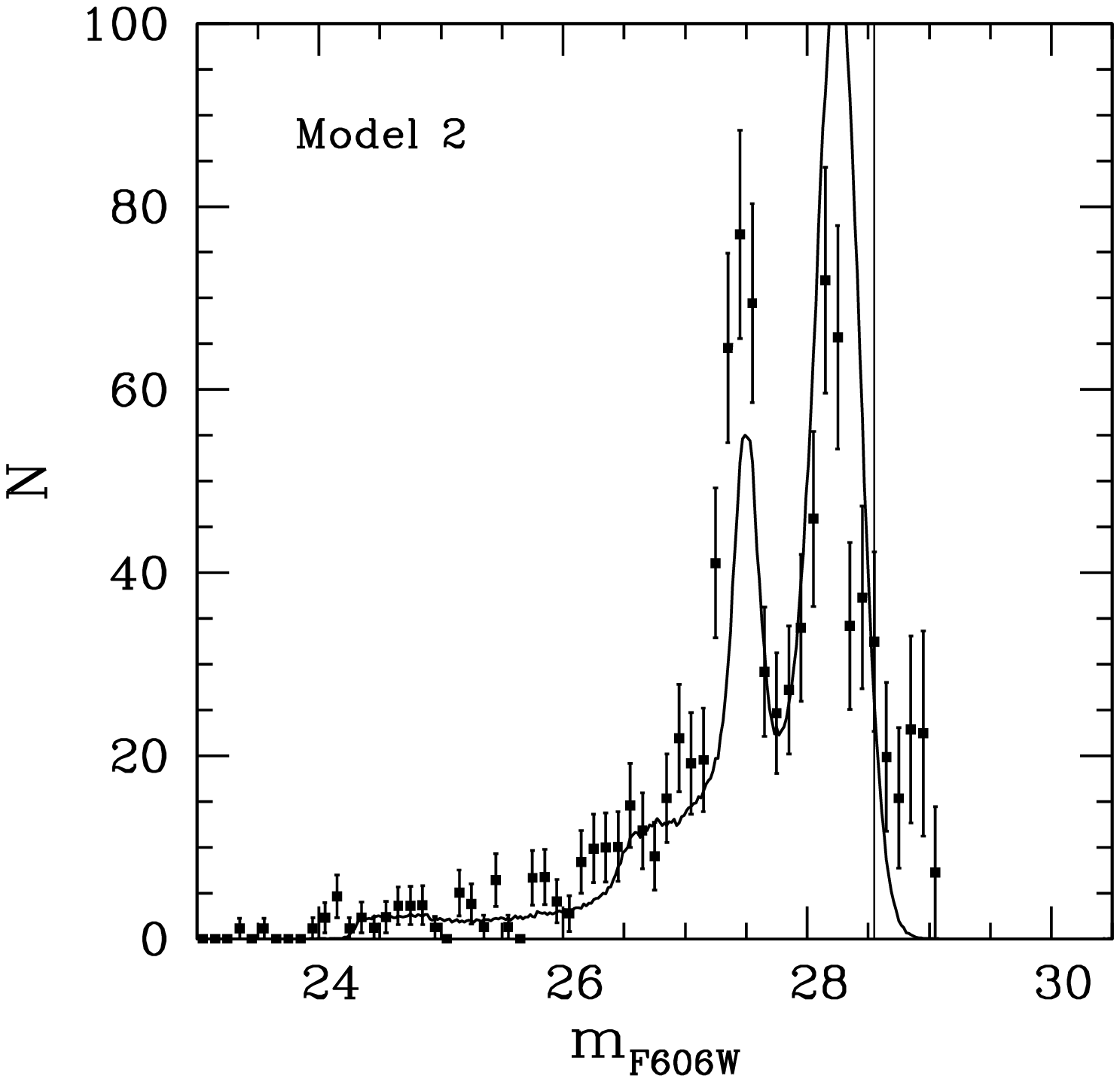}
\includegraphics{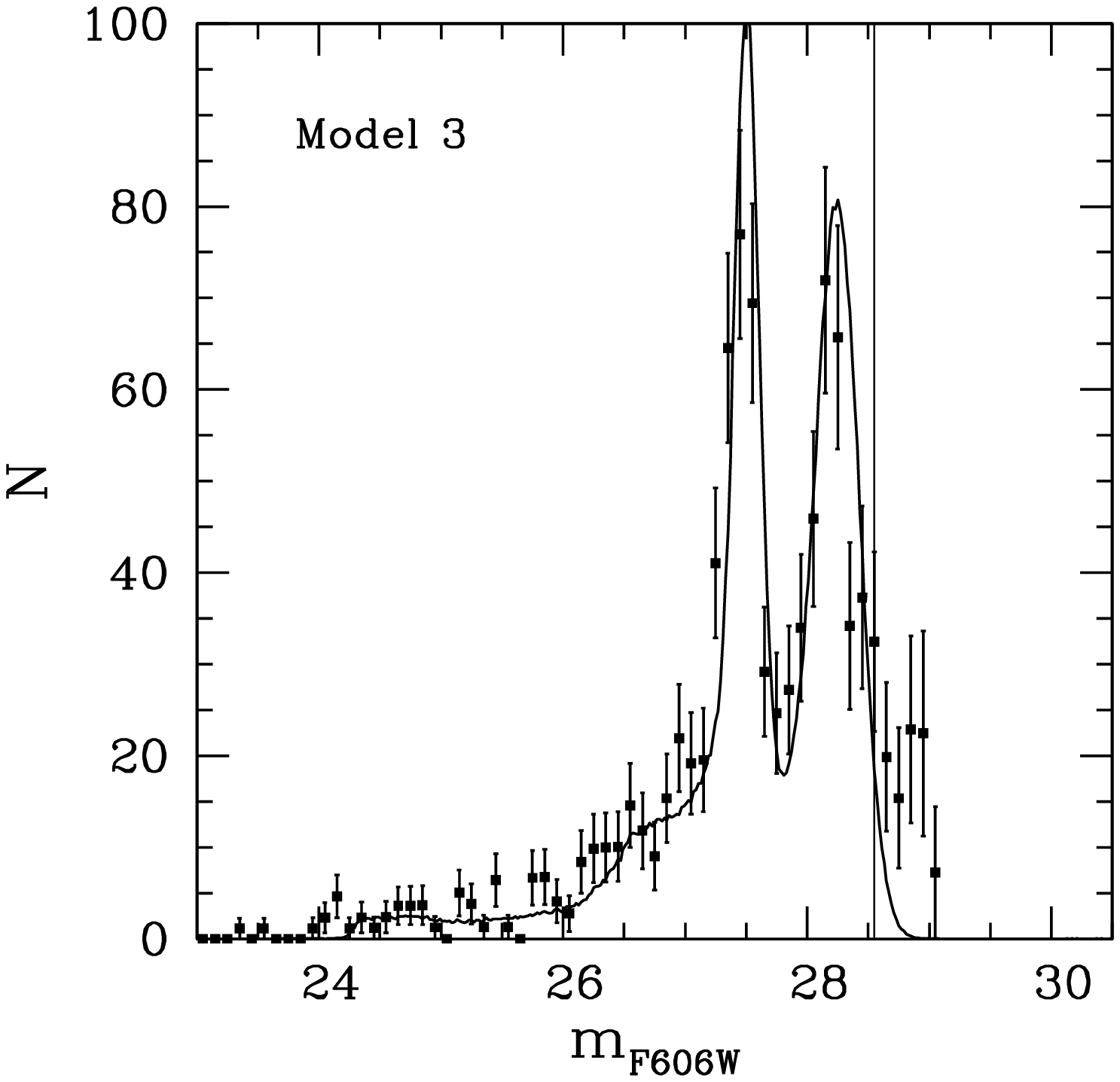} 
\includegraphics{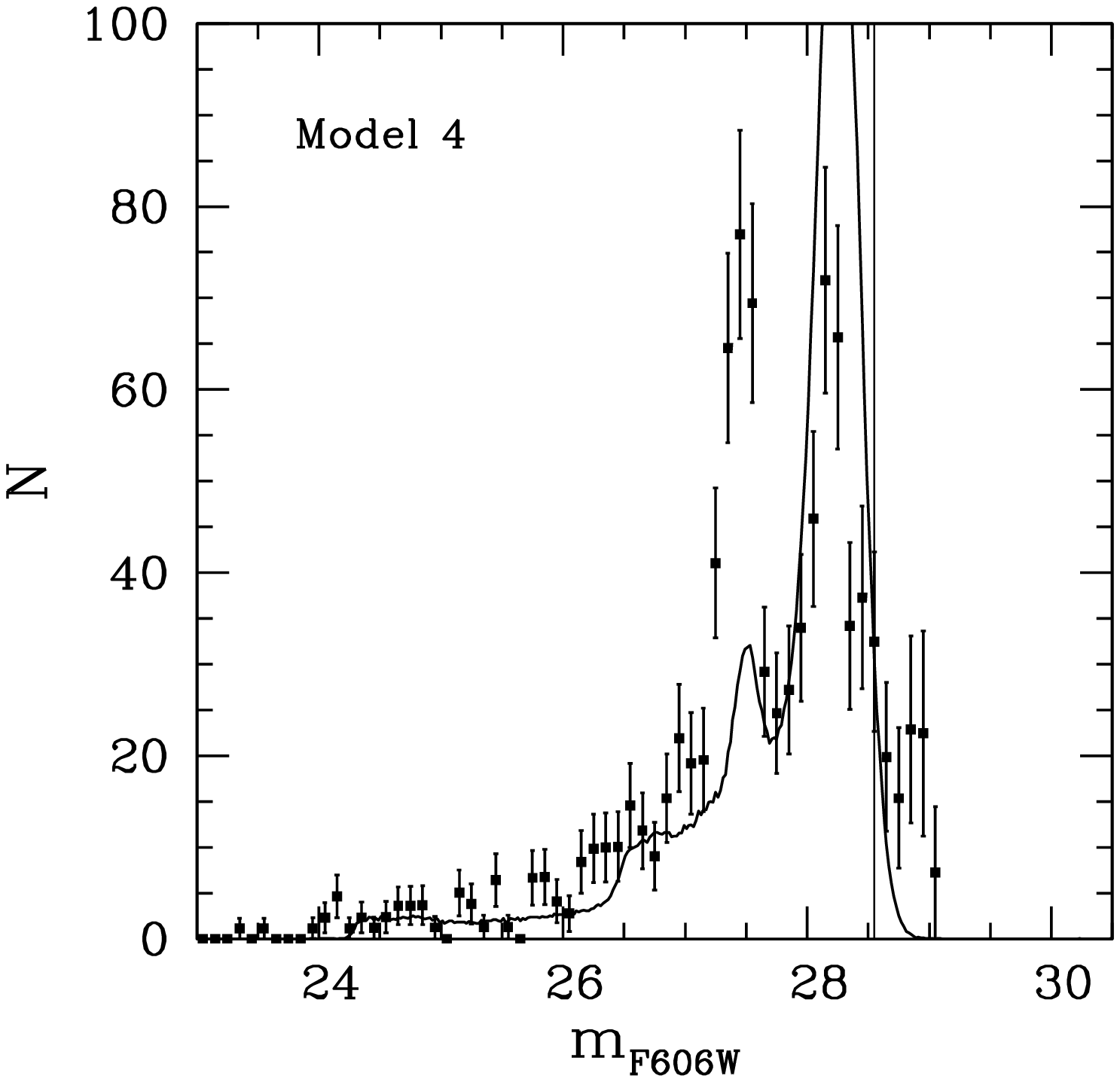}  
\caption{White-dwarf luminosity functions for several distributions of
         secondary masses  of the  progenitor binary system,  see text
         for details.}
\label{m1m2}
\end{figure*}

Now  that we  have established  that single  helium-core  white dwarfs
cannot explain the bright peak of the white-dwarf luminosity function,
a natural  question arises. Namely,  which is the maximum  fraction of
these  white  dwarfs compatible  with  observations?   To answer  this
question we performed additional simulations using our fiducial model,
in which  a population of  unresolved detached binaries is  adopted --
see Sect.~\ref{MC}  -- and we added  a small fraction  $f_{\rm He}$ of
helium-core white dwarfs, keeping  the fraction of binary white dwarfs
of the cluster constant.  For illustrative purposes, the bottom panels
of  Fig.~\ref{helio}  show the  case  in  which  $f_{\rm He}=0.3$  and
$f_{\rm  bin}=0.54$  are adopted.   A  $\chi^2$  test  shows that  the
maximum fraction of single massive helium-core white dwarfs allowed by
the observations is $f_{\rm He}=0.05$.

\subsection{The properties of the binary population}
\label{binary}

If the bright peak of  the white-dwarf luminosity function of NGC~6791
can  only be  explained by  a  population of  unresolved binary  white
dwarfs,  their properties  can  be constrained  using the  white-dwarf
luminosity  function.   There  are  other  clusters  (either  open  or
globular)  for  which  we  have observational  white-dwarf  luminosity
functions  -- M67  (Richer et  al.   1998), NGC~2099  (Kalirai et  al.
2001), NGC~188 (Andreuzzi et al.  2002), M4 (Hansen et al.  2004), and
NGC~6397 (Hansen  et al.  2007;  Richer et al.   2008) -- and  none of
them  shows a secondary  peak in  the white-dwarf  luminosity function
with  the  characteristics of  that  of  NGC~6791.   Perhaps the  best
studied of these  clusters is NGC~6397.  This cluster  has been imaged
down to  very faint  luminosities allowing Hansen  et al.   (2007) and
Richer  et al.   (2008) to  obtain a  reliable  white-dwarf luminosity
function.   The white-dwarf luminosity  function of  this very  old --
about  12  Gyr   (di  Criscenzo  et  al.   2010)   --  and  metal-poor
([Fe/H]$\simeq  -2.03\pm 0.05$)  globular  cluster does  not show  any
evidence  of a  large population  of  binary white  dwarfs.  There  is
another open  cluster for  which a large  number of binaries  has been
found, M67 (Richer et al. 1998).   The age of this cluster is 4.0 Gyr,
and  its  metallicity  is  almost  solar,  [Fe/H]$\simeq  -0.04$,  but
unfortunately  the   small  number  of  white   dwarfs  with  reliable
photometry  does  not  allow  to  perform  a  thorough  study  of  its
population of binary white dwarfs.  All in all, NGC~6791 offers us the
unique  possibility  to study  the  properties  of  the population  of
binaries in an open, very  old and metal-rich cluster.  In particular,
we study  how the white-dwarf luminosity function  allows to constrain
the distribution of secondary  masses.  However, before performing our
analysis  we note that  blending may  have the  same effect  than true
unresolved binaries,  although in the case  of an open  cluster, it is
expected to be less frequent, owing  to the lower density of stars. To
quantify this, we distributed 900 synthetic white dwarfs (the observed
number of  stars) in  the field  of view of  the HST  CCD ($4052\times
4052$   pixels)   and  we   evaluated   the   probability  of   chance
superposition.  We found that this  probability is $\sim 0.8$\% if the
distance necessary  to resolve two  stars is $\sim 10$  pixels.  Thus,
for the case  of NGC~6791 this possibility is  quite unlikely, and the
unresolved binary white dwarfs are most probably real systems.

We used four different models for the distribution of secondary masses
in the progenitor  bynary system, under the assumption  -- the same as
in Bedin et al.  (2008b) -- that binary white dwarfs are produced by a
primordial binary system.  Our first distribution is that already used
by  Bedin  et al.   (2008b),  $n(q)=0.0$  for  $q<0.5$ and  $n(q)=1.0$
otherwise, where  $q=M_2/M_1$, being $M_1$  and $M_2$ the mass  of the
primary  and  of  the  secondary,  respectively.   We  refer  to  this
distribution  as   model  1.   In   model  2  we   assume  $n(q)=1.0$,
independently of the mass ratio.   For model 3 we adopted $n(q)\propto
q$.  Finally, for our last  set of simulations, corresponding to model
4, we adopted $n(q)\propto q^{-1}$.

\begin{figure*}
\vspace{12cm}    
\includegraphics{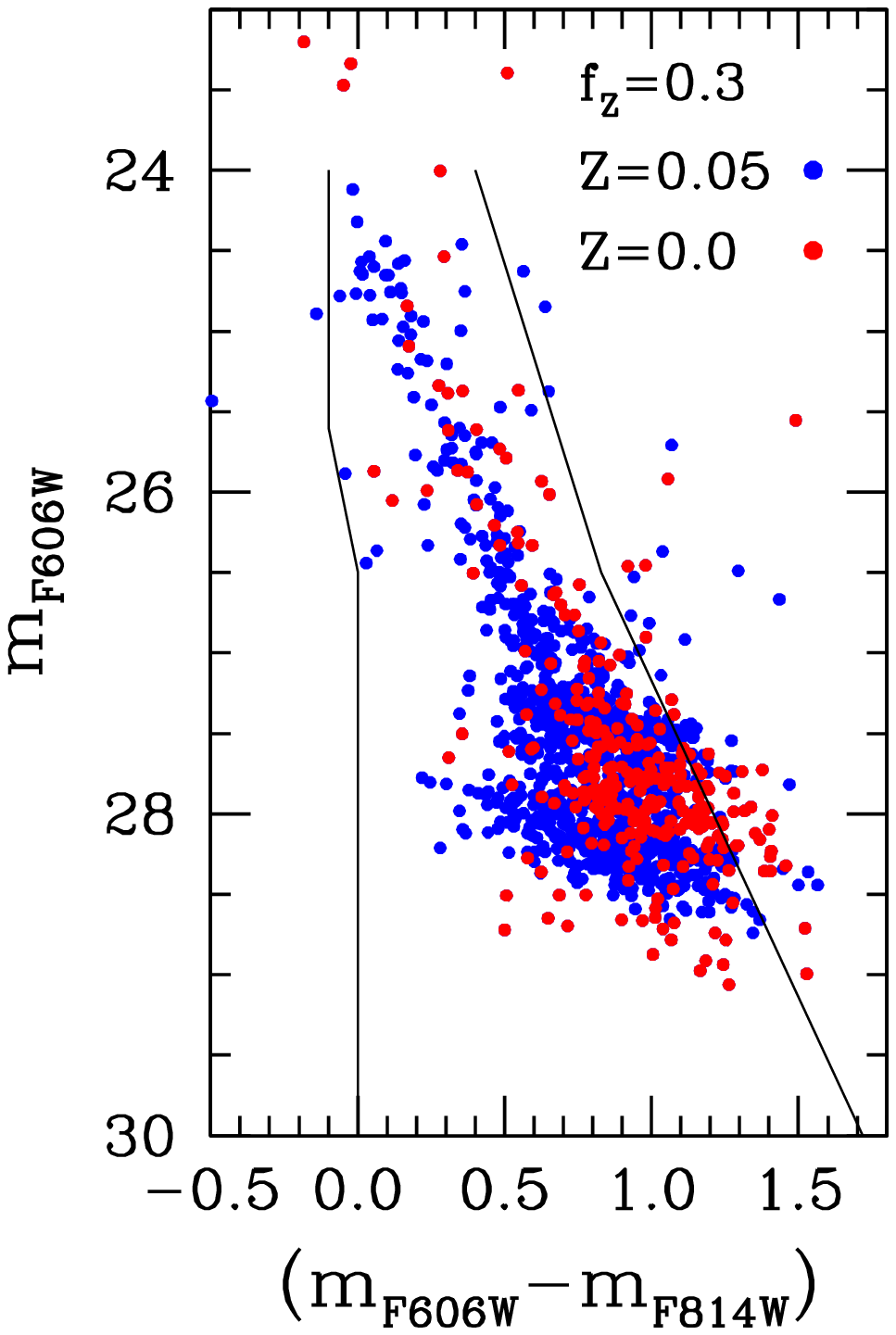}
\includegraphics{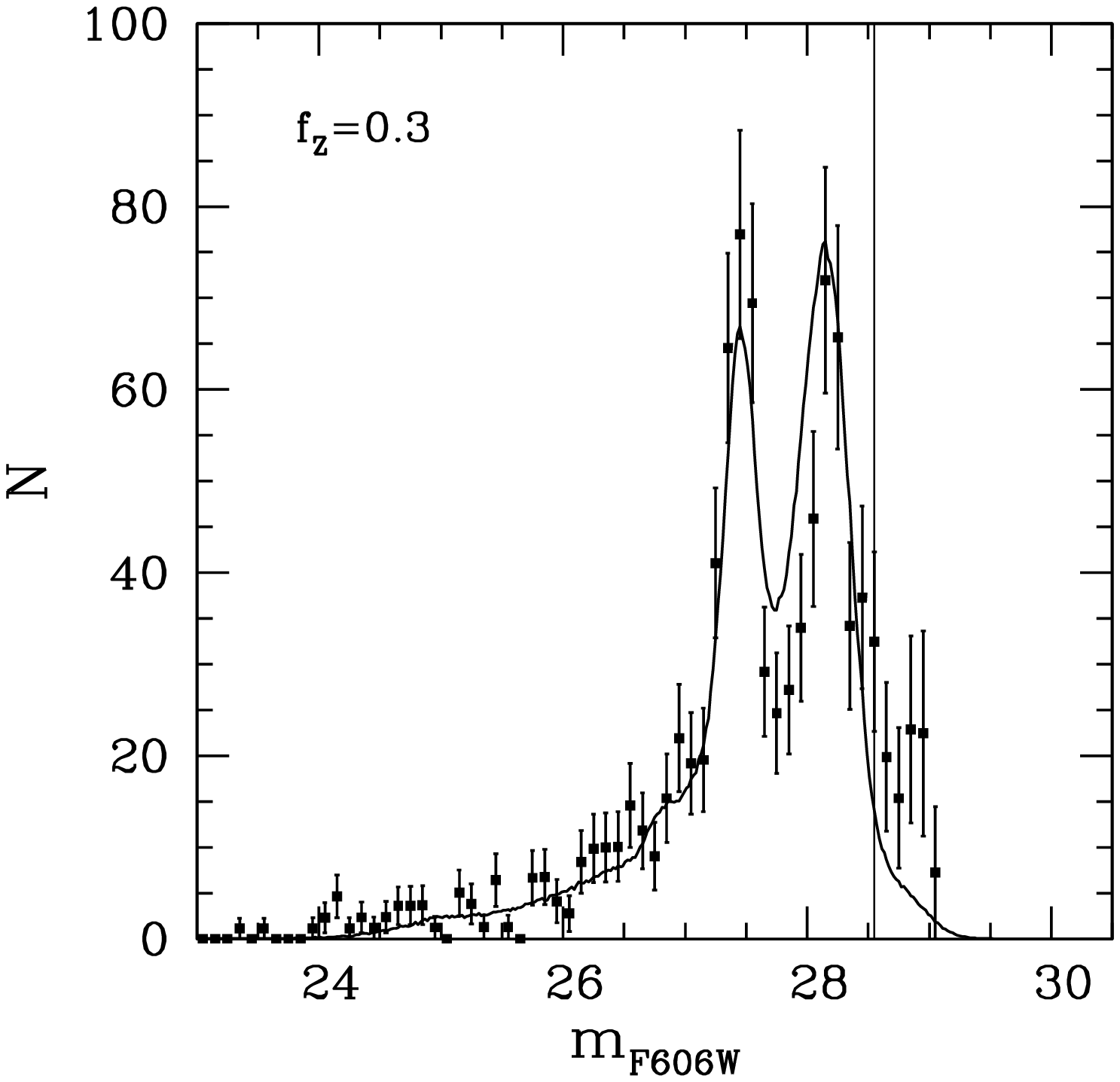}
\includegraphics{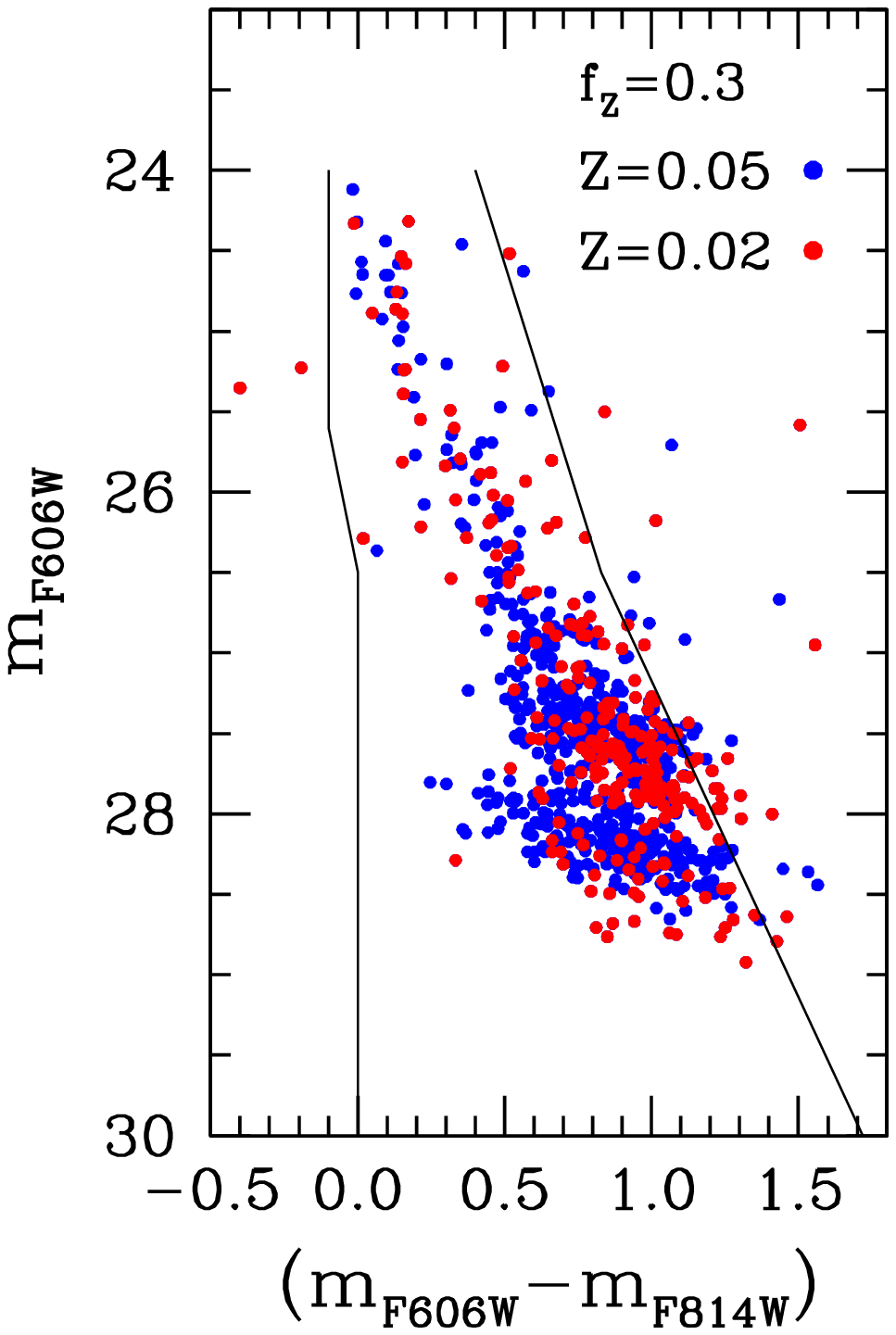} 
\includegraphics{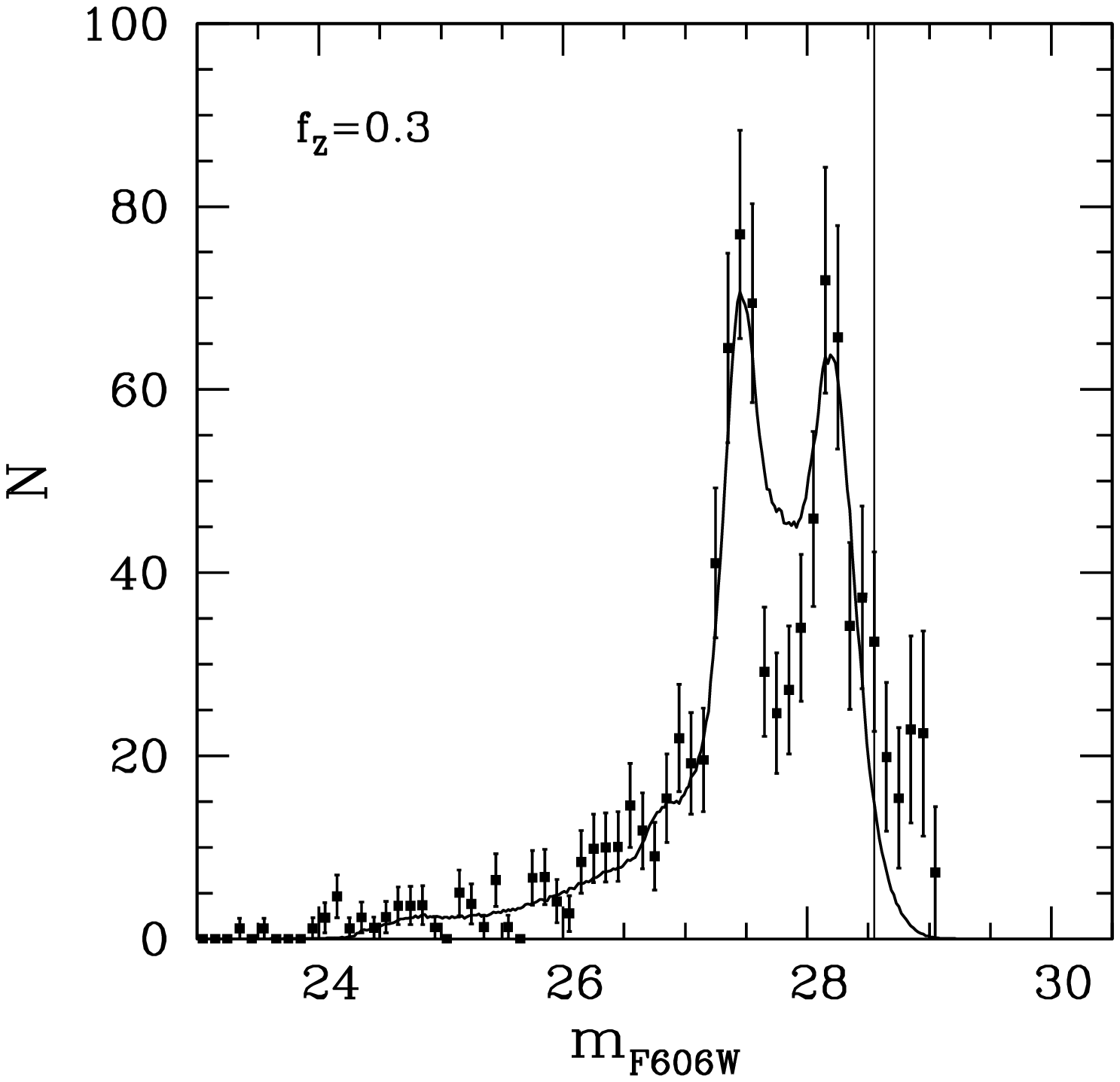}  
\caption{Color-color   diagrams  (left   panels)   of  the   simulated
         subpopulations  of white  dwarfs with  metal-rich progenitors
         (blue circles) and  metal-poor progenitors (red circles), for
         two metallicites  of the subpopulations,  $Z=0.0$ (top panel)
         and  $Z=0.02$ (bottom)  panel.   In both  cases  we show  the
         results for  a fraction  of the subpopulation $f_Z=0.3$.  The
         resulting white-dwarf  luminosity functions (solid lines) are
         compared  to the  observational  one (squares)  in the  right
         panels. See  the online  edition of the  journal for  a color
         version of this plot.}
\label{Z0}
\end{figure*}

We display the results of  this set of simulations in Fig.~\ref{m1m2}.
Evidently,  the corresponding  white-dwarf  luminosity functions  show
dramatic differences.   The distribution of secondary  masses of Bedin
et al.   (2008b), top left panel, perfectly  matches the observational
white-dwarf luminosity function of NGC~6791.  When model 2 is adopted,
the secondary peak of the simulated luminosity function does not match
the observational data, and the amplitude of the faintest peak is very
much increased.   It might be  argued that this incongruence  could be
fixed  by simply  changing the  fraction of  binary white  dwarfs, and
indeed this  could be done,  but then one  would need a  present total
percentage  of  binary  stars  well  above  60\%,  which  is  probably
unrealistic.  Thus, we conclude  that a flat distribution of secondary
masses  can be discarded.   When the  third distribution  of secondary
masses  is used,  we  obtain a  good  fit to  the observational  data,
although the  quality of the fit  is not as  good as that of  model 1.
This is  not surprising, since both distributions  of secondary masses
increase  for increasing  values  of  $q$. Finally,  when  model 4  is
employed,  the simulated  white-dwarf luminosity  function  is totally
incompatible  with the  observational data.   The same  arguments used
when discussing the flat  distribution of secondary masses apply here,
and thus  we can safely  discard this distribution.  We  conclude that
most likely  only distributions of  secondary masses that  increase as
the  mass ratio  of the  two components  of the  binary  increases are
compatible with the existing observational data for NGC~6791.

\subsection{Identification of cluster subpopulations: a test case}
\label{subpopulation}

Our  understanding   of  Galactic  open  and   globular  clusters  has
dramatically changed  in recent years  owing to the wealth  of precise
photometric  data.  This  has allowed  us  to unveil  the presence  of
multiple  main sequences or  subgiant branches  -- see,  for instance,
Piotto et al.  (2007) and Milone et al.  (2008) -- in several globular
clusters. Nowadays there is a handful of Galactic globular clusters in
which the  presence of several subpopulations is  notorious.  The best
known  of these  clusters is  $\omega$~Cen, for  which  four different
metallicity  regimes have  been  so far  identified  (Calamida et  al.
2009).  However, $\omega$~Cen is  not the only example.  For instance,
Piotto et al.   (2007) have convincingly shown that  the main sequence
of   the   globular   cluster   NGC~2808   contains   three   distinct
subpopulations,  while Milone  et al.   (2008) have  demonstrated that
NGC~1851  hosts  a  double  subgiant  branch,  and  di  Criscienzo  et
al.  (2010) have shown  that NGC~6397  may contain  a large  number of
second-generation stars.   Thus, the presence  of multiple populations
in globular clusters is not an infrequent phenomenon.

To the best  of our knowledge there is no  evidence for the occurrence
of this phenomenon in old open clusters. NGC~6791 is particularly well
suited  to  study  this  possibility.   Firstly, it  is  very  old,  a
characteristic shared  with globular clusters.   Secondly, NGC~6791 is
extremely metal-rich.   The origin of this  metallicity enhancement is
still  unknown,  but arguably  could  be due  to  the  existence of  a
previous generation of metal-poor stars.  Nevertheless, we stress that
in other clusters these subpopulations are observed as multiple values
of [Fe/H] (and also helium),  but a recent spectroscopical analysis of
NGC~6791 by  Origlia et al.  (2006)  shows that there is  no spread in
[Fe/H],  and no  spread in  carbon  or oxygen,  which are  two of  the
elements   involved  in   the  related   subpopulations.    Also,  the
well-defined  red   giant  branch  argues   against  this  hypothesis.
However,  there are  other clusters,  of  which NGC~6397  is the  best
example,  which show  remarkably clean  color-magnitude  diagrams with
very tight main sequences and compact blue horizontal branches -- that
is, with no  obvious photometric signs of multiple  populations -- for
which subpopulations have already been identified (Lind et al.  2010).
Finally,  nobody  has  yet  explored  the  possibility  of  using  the
white-dwarf  luminosity  function   to  put  constraints  on  multiple
subpopulations in clusters, although  Prada Moroni \& Straniero (2007)
already  noted  that  the  white  dwarf  isochrones  are  considerably
affected by metallicity variations.   Thus, it is interesting to carry
out  this sensitivity  study  taking advantage  of the  well-populated
white dwarf color-magnitude diagram, to test whether just modeling the
white  dwarf population  can  exclude the  presence of  subpopulations
generated  by progenitors with  a metallicity  different from  the one
measured spectroscopically.

To perform this study, we  first considered varying fractions $f_Z$ of
an extreme  subpopulation with zero metallicity.   The pre-white dwarf
lifetimes  were   taken  from  Marigo  et  al.    (2001),  whilst  the
carbon-oxygen  stratification  in  the  white dwarf  models  was  kept
unchanged, given  the small effect of the  progenitor initial chemical
composition on  the final carbon-oxygen profiles and  cooling times at
fixed white  dwarf mass -- see,  for instance, Salaris  et al.  (2010)
and Renedo et al.  (2010).   We also neglected the delay introduced by
$^{22}$Ne sedimentation,  to account  for the negiglible  abundance of
this element in  $Z=0$ models.  The reason for  this is the following.
$^{22}$Ne is produced during the  helium burning phase by the chain of
reactions  $^{14}$N$(\alpha,  \gamma)^{18}$F$(\beta^+)^{18}$O$(\alpha,
\gamma)^{22}$Ne.   The  net effect  is  to  transform essentially  all
$^{14}$N  into $^{22}$Ne.   In the  extreme case  of $Z=0$  stars some
$^{14}$N is produced  when the CNO cycle is  activated by the $^{12}$C
produced  by  3$\alpha$ reactions  ignited  by  the high  temperatures
during  the main  sequence phase  of  metal free  stars. However,  its
abundance mass fraction is $\la 10^{-9}$ (Weiss et al.  2000).

\begin{figure*}[t]
\vspace{6cm}    
\includegraphics{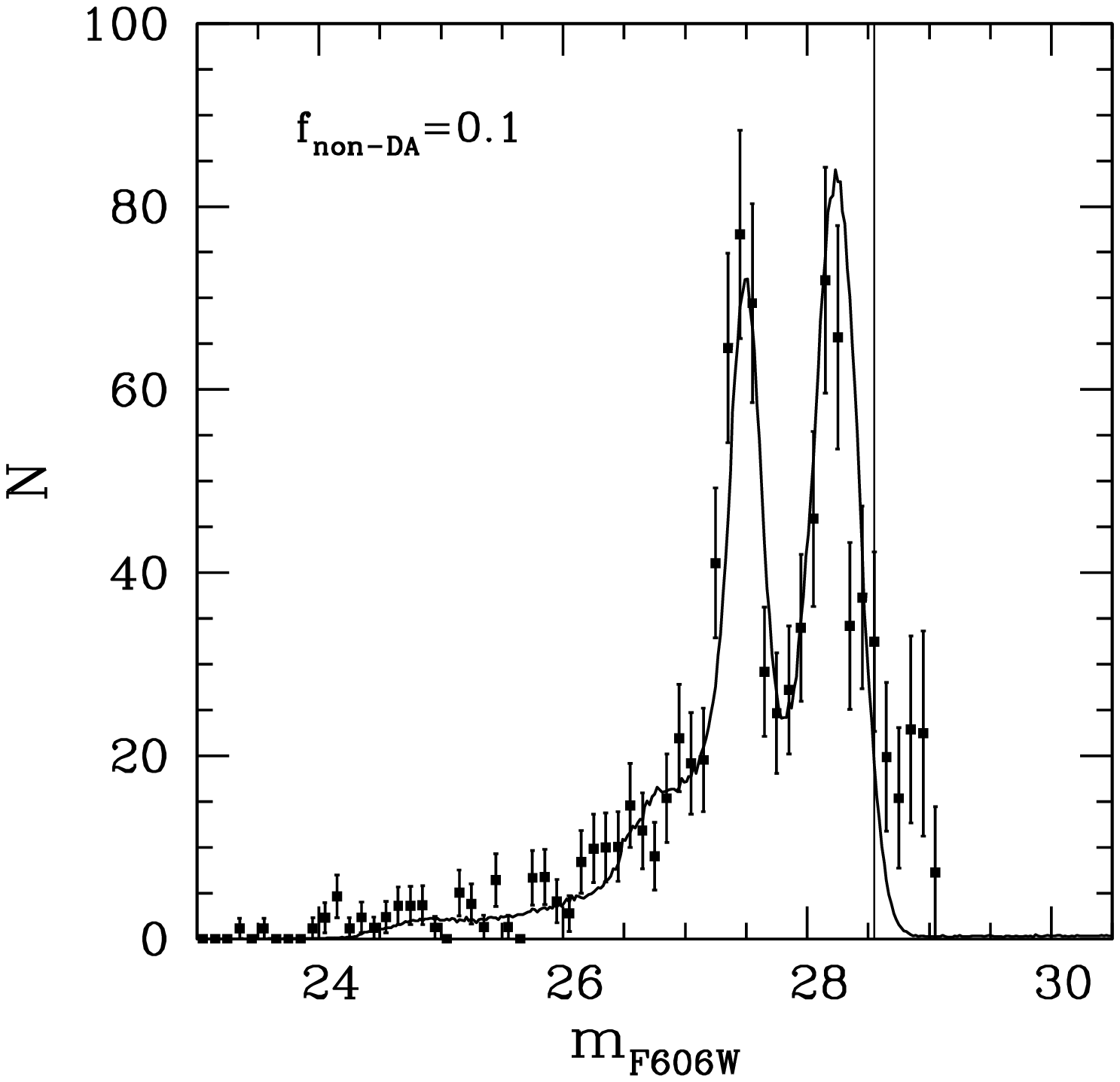}
\includegraphics{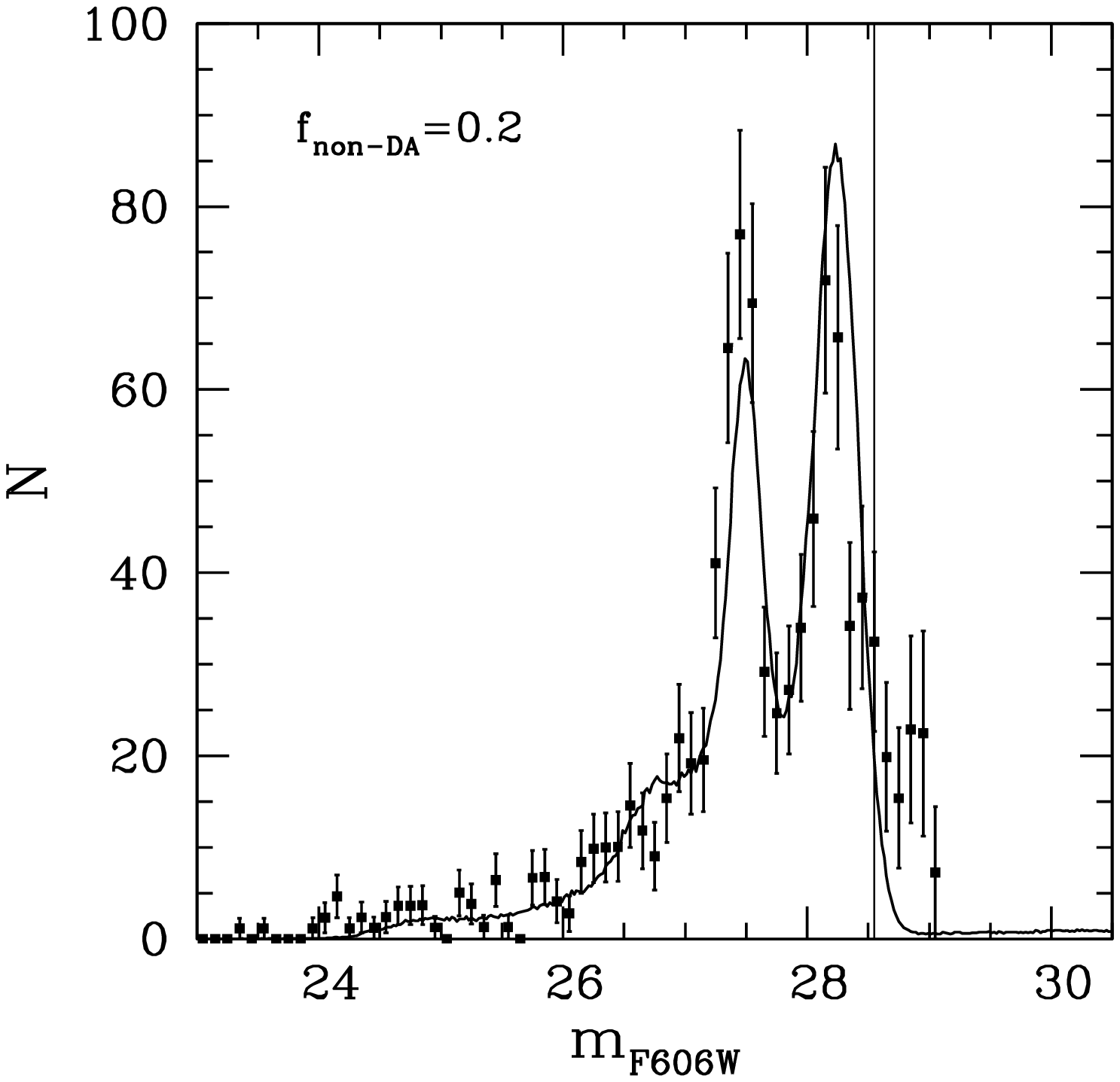}
\caption{Simulated  luminosity functions  for  different fractions  of
         non-DA white dwarfs, as shown in the corresponding panel.}
\label{noDA}
\end{figure*}

The results of  our simulations are shown in  Fig. \ref{Z0}.  We begin
by discussing  the color-magnitude diagram shown in  upper left panel.
Obviously, white  dwarfs resulting from  metal-poor progenitors detach
from the bulk of the  population, and several of these synthetic white
dwarfs can  be found  below the well-defined  cut-off of  the observed
cooling sequence, as expected, because  the lack of $^{22}$Ne causes a
faster  cooling.  The  upper  right  panel of  this  figure shows  the
corresponding  luminosity function.   The overall  agreement  with the
observed luminosity function is poor, especially in the region between
the two  peaks.  An increase of  the binary fraction  to reproduce the
bright  peak better  would  not  improve the  modeling  of the  region
between the peaks.   The natural question to address  is then which is
the maximum fraction of metal-poor white dwarf progenitors that can be
accommodated within the observational white-dwarf luminosity function?
To this  purpose we have  computed synthetic white dwarf  samples with
decreasing fractions of metal-poor  progenitors, and we ran a $\chi^2$
test.   We  found that  the  maximum  allowed  fraction of  metal-poor
progenitors  is  $f_Z=0.12$.   Obviously,  the  assumption  that  this
subpopulation   has   zero  metallicity   is   probably  to   extreme.
Consequently, we repeated the  same calculation for a subpopulation of
solar metallicity --  see the bottom panels of  Fig.~\ref{Z0}. In this
case  the   maximum  fraction  of   solar-metallicity  progenitors  is
$f_Z=0.08$.

\subsection{The fraction of non-DA white dwarfs}
\label{noDAf}

We now focus on the  possibility of determining the fraction of non-DA
white  dwarfs in  NGC~6791.   The spectral  evolution  of white  dwarf
atmospheres is still a  controversial question, and although the ratio
of white dwarfs with pure hydrogen atmosphere versus white dwarfs with
hydrogen-deficient atmospheres is known  for the local field, very few
determinations  exist  for  open  and  globular  clusters.   Moreover,
although for the field white dwarf population the canonical percentage
is  around 80\%,  observations show  that  this ratio  depends on  the
effective  temperature  -- see,  for  instance,  Tremblay \&  Bergeron
(2008)   and   references  therein.    However,   the  only   reliable
determinations for open  clusters are those of Kalirai  et al.  (2005)
for the  rich, young  cluster NGC~2099, and  Rubin et al.   (2008) for
NGC~1039. Kalirai et al.  (2005) found a clear deficit of non-DA white
dwarfs  in NGC~2099,  whereas  Rubin  et al.   (2008)  found that  the
fraction of non-DA white dwarfs  in the open cluster NGC~1039 is $\sim
10\%$,  at most.   Clearly, investigating  the DA  to non-DA  ratio in
another open cluster is therefore of greatest interest.

We  addressed this question  by simulating  the cluster  population of
white dwarfs with an increasing  fraction of non-DA stars.  The non-DA
fractions  adopted here are  $f_{\rm non-DA}=0.0$,  0.1, 0.2  and 0.4,
respectively.  For the  sake of conciseness, we only  show the results
for $f_{\rm non-DA}=0.1$ -- left panel  -- and 0.2 -- right panel.  As
shown  in  Fig.  \ref{noDA}  the  white-dwarf  luminosity function  is
sensitive to  the ratio of  non-DA to DA  white dwarfs.  We  find that
when the fraction  of non-DA white dwarfs is  increased, the agreement
with  the   observational  white-dwarf  luminosity   function  rapidly
degrades.  To be precise, when  the fraction of non-DA white dwarfs is
$f_{\rm  non-DA}=0.1$,  the agreement  is  quite  poor,  and when  the
fraction of non-DA white dwarfs is that of field white dwarfs, $f_{\rm
non-DA}=0.2$, the quality of  the fit to the observational white-dwarf
luminosity function is  unacceptable.  This is because for  the age of
NGC~6791 non-DA  and DA white dwarfs pile-up  at similar luminosities.
As a consequence, adding single  non-DA white dwarfs lowers the height
of the bright peak compared to the faint one.

To quantify which the maximum fraction of non-DA white dwarfs that can
be  accommodated within the  observational errors  is, we  conducted a
$\chi^2$ test, and we found  that for fractions of non-DA white dwarfs
larger than $\sim 0.1$ the probability rapidly drops below $\sim 0.7$,
whereas  for  $f_{\rm  non-DA}=0.0$  the probability  is  $\sim  0.9$.
Consequently,  the fraction  of non-DA  white dwarfs  in  NGC~6791 can
roughly be at most half the  value found for field white dwarfs.  This
result  qualitatively  agrees with  the  findings  of  Kalirai et  al.
(2005), who find that for NGC~2099 this deficit of non-DA white dwarfs
is even  higher.  As a  matter of fact,  Kalirai et al.   (2005) found
that for this cluster all white  dwarfs in their sample were of the DA
type.   Our results  also  point  in the  same  direction, although  a
fraction  of  $\sim  5\%$   is  still  compatible  with  the  observed
white-dwarf  luminosity function  of NGC~6791,  in agreement  with the
findings of Kalirai et al.~(2007).

Finally,  we considered  also  the possibility  that  the fraction  of
non-DA  white dwarfs  changes  with the  effective temperature,  which
occurs with  field white dwarfs.   In particular, we assumed  that for
effective temperatures  higher than  6\,000~K, the fraction  of non-DA
white dwarfs is $f_{\rm non-DA}=0.2$ and for temperatures ranging from
5\,000~K   to  6\,000~K,   $f_{\rm  non-DA}=0.0$,   as   suggested  by
observations of  low-luminosity field white  dwarfs.  In this  case we
find   that  the   simulated  white-dwarf   luminosity   function  and
color-magnitude  diagram are very  similar to  those in  which $f_{\rm
non-DA}=0.0$,  and  thus  agree   very  well  with  the  observational
luminosity function of NGC~6791. Nevertheless, it is worth noting that
using this prescription, the  fraction of non-DA white dwarfs expected
in the  cluster would  be about  $6\%$.  Thus, on  the basis  of these
simulations  it  cannot be  discarded  that  this  cluster could  have
originally produced a large percentage  of non-DA white dwarfs, but at
the  present  age  of  the  cluster,  most of  them  could  have  been
transformed into DA white dwarfs as a result of accretion episodes.

These  conclusions clearly depend  on the  assumed fraction  of binary
white dwarfs  that populate the  cluster. The fraction of  white dwarf
binaries  necessary  to explain  the  bright  peak  of the  luminosity
function (in  absence of non-DA  objects) requires that about  54\% of
the objects in NGC~6791 be  binaries.  An increased white dwarf binary
percentage  (75\%) can  in principle  accommodate a  higher percentage
(20\%)  of non-DA  objects by  increasing the  relative height  of the
bright peak in the synthetic sample, compared to the faint one, but it
seems  unrealistic  to  accept  such  a large  percentage  of  cluster
binaries, hence non-DA white dwarfs.


\section{Conclusions}
\label{concl}

In this paper we have investigated several important properties of the
stellar  population  hosted  by  the  very  old  (8  Gyr),  metal-rich
([Fe/H]$\simeq  0.4$) open  cluster NGC~6791.   This cluster  has been
imaged  below the  luminosity of  the termination  of its  white dwarf
cooling  sequence   (Bedin  et  al.   2005,   2008a).   The  resulting
white-dwarf luminosity  function enables us not only  to determine the
cluster  age (Garc\'\i  a--Berro et  al.  2010),  but  other important
properties as  well.  Among  these, we mention  the properties  of the
population of unresolved binary white dwarfs, the existence of cluster
subpopulations, and the fraction of non-DA white dwarfs.

The origin of  the bright peak of the  white-dwarf luminosity function
was  investigated,   exploring  in  detail   the  alternative  massive
helium-core white  dwarf scenario.  Our  conclusion is that  this peak
cannot  be attributed  to  a population  of  single helium-core  white
dwarfs.  The  more realistic possibility left to  explain this feature
is  a  population  of  unresolved  binary  white  dwarfs.   This  huge
population of unresolved  binary white dwarfs has allowed  us to study
the properties of the parent  population.  To this purpose, we studied
the properties of  the distribution of secondary masses  in the binary
progenitor  system  and  its  effects in  the  white-dwarf  luminosity
function.   Specifically, we  tested four  different  distributions of
secondary masses and  we found that only those  distributions that are
monotonically increasing  with the mass ratio are  consistent with the
observational  data.  Additionally, as  a test  case, we  verified the
ability of the white-dwarf luminosity function to assess the existence
of subpopulations  within a  stellar system.  We  have found  that the
presence of a $Z=0$ subpopulation is inconsistent with the white-dwarf
luminosity function,  the maximum fraction  allowed by the  data being
12\%.  If the metallicity of the subpopulation is solar, this fraction
is 8\%.

Finally, we  found that  the fraction of  non-DA white dwarfs  in this
cluster is  unusually small,  on the  order of 6\%  at most,  and much
smaller than  the corresponding one  for field white dwarfs,  which is
$\sim 20\%$. This shortage of  non-DA white dwarfs is a characteristic
shared with  another open cluster, NGC~2099.  However,  the deficit of
non-DA white dwarfs is even higher in the case of NGC~2099, given that
for this  cluster recent exhaustive observations have  found no single
white dwarf of the non-DA type (Kalirai et al.  2005).


\begin{acknowledgements}
This   research   was   supported    by   AGAUR,   by   MCINN   grants
AYA2008--04211--C02--01  and  AYA08-1839/ESP,  by the  European  Union
FEDER funds, by the  ESF EUROGENESIS project (grant EUI2009-04167), by
AGENCIA: Programa de Modernizaci\'on Tecnol\'ogica BID 1728/OC-AR, and
by PIP 2008-00940 from CONICET.   LGA also acknowledges a PIV grant of
the AGAUR of the Generalitat  de Catalunya. We also thank our referee,
O. Straniero, for useful comments and suggestions.
\end{acknowledgements}


\end{document}